\documentstyle[12pt]{article}
\newlength{\pubnumber} 

\catcode`\@=11
\def\section{\@startsection{section}{1}{\z@}{3.5ex plus 1ex minus .2ex}
 {2.3ex plus .2ex}{\large\bf}}
\def\subsection{\@startsection{subsection}{2}{\z@}{2.3ex plus .2ex}
 {2.3ex plus .2ex}{\bf}}

\begin{document}

\begin{titlepage}
\samepage{
\setcounter{page}{1}
\rightline{McGill/93-18}
\rightline{\tt hep-th/9306156}
\rightline{June 1993}
\vfill
\begin{center}
 {\Large \bf Recent Developments\\
 in Fractional Superstrings\footnote{
   Talk presented at {\it SUSY-93:  International Workshop
   on Supersymmetry and Unification of Fundamental Interactions},
   held at Northeastern University, Boston, MA, 29 March -- 1 April 1993;
   to appear in Proceedings published by World Scientific.}\\}
\vfill
 {\large Keith R. Dienes\footnote{E-mail address:
dien@hep.physics.mcgill.ca}\\}
\vspace{.10in}
 {\it  Department of Physics, McGill University\\
3600 University St., Montr\'eal, Qu\'ebec~H3A-2T8~~Canada\\}
\end{center}
\vfill
\begin{abstract}
  {\rm Fractional superstrings experience new types of ``internal
  projections'' which alter or deform their underlying worldsheet conformal
  field theories.  In this talk I summarize
  some recent results concerning both the worldsheet theory
  which remains after the internal projections have acted, and
  the spacetime statistics properties of its various sectors.}
\end{abstract}
\vfill}
\end{titlepage}

\setcounter{footnote}{0}
%========================================================================
%           DEFINITIONS and HYPHENATIONS
%========================================================================
\def\beq{\begin{equation}}
\def\eeq{\end{equation}}
\def\beqn{\begin{eqnarray}}
\def\eeqn{\end{eqnarray}}
\def\calZ{{\cal Z}}
\def\half{{\textstyle {1\over2}}}
\def\bone{{\bf 1}}
\def\ie{{\it i.e.}}
\def\eg{{\it e.g.}}
 \font\cmss=cmss10 \font\cmsss=cmss10 at 7pt
\def\IZ{\relax\ifmmode\mathchoice
 {\hbox{\cmss Z\kern-.4em Z}}{\hbox{\cmss Z\kern-.4em Z}}
 {\lower.9pt\hbox{\cmsss Z\kern-.4em Z}}
 {\lower1.2pt\hbox{\cmsss Z\kern-.4em Z}}\else{\cmss Z\kern-.4em Z}\fi}

\hyphenation{pa-ra-fer-mion pa-ra-fer-mion-ic pa-ra-fer-mions }
\hyphenation{su-per-string frac-tion-ally su-per-re-pa-ra-met-ri-za-tion}
\hyphenation{su-per-sym-met-ric frac-tion-ally-su-per-sym-met-ric}
\hyphenation{space-time-super-sym-met-ric fer-mi-on}
\hyphenation{Ne-veu-Schwarz-like Ramond-like}
%=====================================================================

%============================== SECTION 1 ============================

\setcounter{footnote}{0}
\section{Introduction}

Over the past two years there has been considerable
activity in a possible new class of string theories
known as {\it fractional superstrings} [1--10]:
these are non-trivial generalizations of superstrings and heterotic strings,
and have the important property that
their critical spacetime dimensions are less than ten.
This reduction in the critical dimension
is accomplished by replacing the worldsheet supersymmetry
of the traditional superstring or heterotic string
by a $K$-fractional supersymmetry:
such symmetries relate worldsheet bosons not to worldsheet fermions,
but rather to worldsheet $\IZ_K$ parafermions $\epsilon$
of fractional spin $2/(K+2)$.
One then finds that the corresponding
critical spacetime dimension of the theory is given by
\beq
   D_c ~=~ 2 ~+~ {{16}\over K} ~,~~~~K\geq 2~.
\label{kcritdimensions}
\eeq
Thus while the choice $K=2$ reproduces the ordinary $D_c=10$ superstring
(with $\IZ_2$ ``parafermions'' reducing to ordinary Majorana fermions),
the choices $K=4,8,$ and $16$ yield new theories with $D_c=6,4,$ and $3$
respectively.

For $K>2$, the $\IZ_K$ parafermion conformal field theory (CFT)
is non-linear:  the appearance of fractional-spin fields
implies that their operator-product expansions contain cuts rather than poles,
and indeed these fields have non-trivial (and often non-abelian)
braiding relations.
It is primarily due to such complications on the worldsheet that
fractional superstrings appear to exhibit qualitatively new features
in spacetime, as compared to the usual superstrings and heterotic strings.
Understanding these new features
is thus of paramount importance, not only for demonstrating
the internal consistency of the fractional superstring,
but also as a means of shedding further light on
the general but as yet poorly understood relationship between worldsheet
string symmetries and spacetime physics.

While the low-lying states of the fractional superstring
are closely analogous those of the ordinary superstring,
the non-linearity of the fractional-superstring
worldsheet theory becomes manifest at higher mass levels.
In particular, two fundamentally new features emerge
whose spacetime interpretations have thus far remained unresolved.
The first is the appearance of new massive sectors
which are not of the standard Ramond or Neveu-Schwarz (NS) variety,
and which contain spacetime particles whose physical roles (and
spacetime statistics properties) are unclear.
The second is the appearance of new so-called ``internal projections''
which, unlike the traditional GSO projection, appear to change or deform
the parafermionic conformal field theories
upon which the fractional superstring is built, leaving behind
worldsheet CFT's whose properties are as yet unknown.
We shall here provide a short review of recent
developments in these areas, referring the reader to
Ref.~\cite{Review} for a non-technical overview
of fractional superstrings, to
the original papers (Refs.~[1--4]) for more information
concerning the basic ideas behind fractional superstrings,
and to Refs.~\cite{AD,D} for further details
concerning these new results.

%========================================================================
\setcounter{footnote}{0}
\section{Massive Sectors}

In order to specify the sense in which the fractional superstring
contains new types of sectors, it proves
instructive to recall the case of the ordinary superstring in $D=10$.
The underlying light-cone worldsheet CFT
of the usual superstring has central charge $c=12$, and consists of a tensor
product of eight free bosons and eight Ising models, one copy of each
per transverse spacetime dimension.
Each of these Ising-model CFT's contains three fields:  the identity $\bone$,
the Majorana fermion $\psi$, and the spin-field $\sigma$.
There are thus a variety of combinations of CFT sectors which
could potentially contribute states to the superstring spacetime spectrum.
However, as is well-known, only five sectors
actually contribute to the spectrum:
these are the four NS sectors with positive $G$-parity
$\bone^7\psi, \bone^5\psi^3, \bone^3\psi^5$, and $\bone\psi^7$,
and the single Ramond sector $\sigma^8$.
Since the spin-field $\sigma$ introduces a cut on the worldsheet
which changes the boundary conditions of the worldsheet fermion $\psi$,
we find in the usual way that
states in the NS sectors are spacetime bosons, and states in the Ramond
sector are spacetime fermions.
The crucial observation, however, is that no
``mixed'' $\bone/\sigma$, $\psi/\sigma$, or $\bone/\psi/\sigma$
combinations contribute to the physical spectrum of
states of the ordinary $D=10$ superstring.

For the more general fractional superstrings with $K>2$, this is no longer
the case:  there are a variety of fundamentally new sectors which contribute
states to the spacetime spectrum and which must therefore be considered.
These can be described as follows.
In analogy to the superstring, the light-cone worldsheet CFT
of the $K$-fractional superstring consists of $D_c-2=16/K$ free bosons
tensored together with $16/K$ copies of the $\IZ_K$ parafermion
theory;  these $\IZ_K$ parafermion theories are fractional-spin generalizations
of the Ising model.
Thus, since each $\IZ_K$ parafermion theory has central charge
$c_K=(2K-2)/(K+2)$, the light-cone
worldsheet CFT of the $K$-fractional superstring
has total central charge $c=48/(K+2)$:
\beq
   K\geq 2:~~~~~~
   \left(c={48\over K+2}\right)~{\rm CFT} ~=~
     \left\lbrace \mathop\otimes_{\mu=1}^{D_c-2=16/K}  X^\mu \right\rbrace
          ~~\otimes~~
     \left\lbrace \mathop\otimes_{\mu=1}^{D_c-2=16/K}  (\IZ_K~{\rm PF})^\mu
     \right\rbrace~.
\label{originalCFT}
\eeq
Like the Ising model, however, each of these $\IZ_K$ parafermion theories
contains a variety of primary fields, and these can be grouped into three
classes:  analogues of the Ising-model fields $\bone$ and $\psi$
(producing spacetime bosonic
states), analogues of $\sigma$ (producing spacetime fermionic states),
and additional parafermionic fields (to be collectively
denoted $\phi$) which have no
analogues in the Ising model.
There are thus sectors of the forms $(\bone,\psi)^{16/K}$
and $(\sigma)^{16/K}$ which are respectively the
fractional-superstring analogues
of the superstring NS and Ramond sectors;  these are the so-called
``$A$-sectors'', and they contain all of the massless states (including,
for example, the supergravity multiplet).
There are, however, two other types of sectors which contribute
to the fractional-superstring spectrum.  The first (the so-called
``$B$-sectors'') all have the equally mixed form
$(\lbrace\bone\rbrace\lbrace\sigma\rbrace)^{8/K}$,
and contain only states with masses
$m^2>0$ ({\it i.e.}, states at the Planck scale).
The second (the so-called ``$C$-sectors'')
instead take the form $(\phi)^{16/K}$, and also contain
only Planck-scale states.
It is these two groups of sectors which are the unusual ``massive sectors''
whose physical properties have thus far remained elusive.

%========================================================================
\setcounter{footnote}{0}
\section{Internal Projections}

The second fundamental issue which has remained unresolved concerns the
appearance of new types of so-called ``internal projections'' which
remove states from the fractional-superstring spectrum.
Whereas the GSO projection in the ordinary superstring
removes only entire {\it towers} of states
(projecting out, for example, the {\it odd}\/ $G$-parity NS sectors
$1^i \psi^{8-i}$ with $i\in 2\IZ$),
these new internal projections
project away only {\it some} of the states in each individual tower,
leaving behind a set of states which therefore cannot
be interpreted as the complete Fock space of the original underlying
worldsheet CFT in Eq.~(\ref{originalCFT}).
On the face of it, this would seem to render the
spacetime spectra of the fractional superstrings hopelessly inconsistent
with any underlying worldsheet-theory interpretation.
Remarkably, however, evidence suggests that the residual states which
survive the internal projections in each tower precisely recombine to
fill out the complete Fock space of a {\it different}\/ underlying
conformal field theory.  Thus, whereas the GSO projection merely removed
certain highest-weight sectors of the worldsheet conformal field theory,
these new internal projections appear to actually {\it change the
underlying conformal field theory itself.}
In fact, since the central charges of the new (post-projection)
CFT's are smaller than those of the
original tensor-product parafermion theories, these internal projections
must clearly remove exponentially large numbers of states from each
of the mass levels of the original Fock space.
Such a drastic projection clearly has no analogue in the
ordinary superstring, and perhaps more closely resembles the BRST projection
which enables unitary minimal models with $c<1$ to be constructed
from free $c=1$ bosons in the Feigin-Fuchs construction.

Verifying that the internal projection in fact leaves behind a
self-consistent Fock space is a difficult task, and to date the evidence
for this has been obtained only through an analysis of the
fractional-superstring partition functions.  Indeed, even the existence of
these internal projections can at present be deduced only through this
partition-function approach, and there does not currently exist any
internal-projection operator constructed out of worldsheet
fields which would enable us to analyze these projections at the level of
individual states.  Therefore the partition functions remain the primary
tool for analyzing these internal projections, and we shall see that this
approach is sufficient to determine the central charges, highest weights,
fusion rules, and characters of the worldsheet conformal field theories
which survive the internal projections in {\it all}\/ of the fractional
superstring sectors.  Moreover, by exploiting the similarity of these
CFT's with those of free compactified bosons, we shall even be able to gain
valuable information concerning the spacetime statistics of the surviving
states.  Actually constructing a suitable representation for this conformal
field theory in terms of worldsheet fields still remains an open question,
however, and we shall discuss some of the difficulties at the end of this
talk.

%=========================================================================
\setcounter{footnote}{0}
\section{Fractional Superstring Partition Functions}

Let us now review the partition-function evidence for these
extra sectors and internal projections.
Given the worldsheet light-cone CFT's of the $K$-fractional superstrings
indicated in Eq.~(\ref{originalCFT}),
it is a straightforward matter to construct their corresponding
modular-invariant one-loop partition functions.
Such partition functions are of course constructed as linear combinations
of products of the characters corresponding to each individual
free coordinate boson plus $\IZ_K$ parafermion system;  these characters
are the so-called ``string functions'', and we
indicate via the schematic notation $\chi_1$,
$\chi_{\sigma}$, and $\chi_{\phi}$
those groups of string functions
which correspond respectively to the Neveu-Schwarz-like, Ramond-like,
or ``other'' parafermion fields of each $\IZ_K$ theory.
It can then be shown [1--3]
that demanding the presence of a massless
sector and the absence of physical tachyons leads to the following
unique partition functions $\calZ_K$ for each relevant value of $K\geq 2$:
\beqn
  \calZ_2&~=&~ ({\rm Im}\,\tau)^{-4\phantom{/1}}\,
     \phantom{\bigl\lbrace}|A_2^b-A_2^f|^2 \nonumber\\
  \calZ_4&~=&~ ({\rm Im}\,\tau)^{-2\phantom{/1}}\,
    \left\lbrace|A_4^b-A_4^f|^2 ~+~ 3\, |B_4|^2\right\rbrace \nonumber\\
  \calZ_8&~=&~ ({\rm Im}\,\tau)^{-1\phantom{/2}}\,
     \left\lbrace|A_8^b-A_8^f|^2 ~+~ |B_8|^2 ~+~2\,|C_8|^2
       \right\rbrace\nonumber\\
  \calZ_{16}&~=&~ ({\rm Im}\,\tau)^{-1/2}\,\left\lbrace
    |A_{16}^b-A_{16}^f|^2 ~+~ |C_{16}|^2 \right\rbrace
\label{partfuncts}
\eeqn
where the expressions $A_K^{b,f}$, $B_K$, and $C_K$
respectively represent the contributions from
the $A$, $B$, and $C$ sectors, and take the forms
\beqn
   &&A_K^{b} ~\sim~ \sum \,g^{(A^b)}_i\,(\chi_{1})^{D_c-2}~,~~~~
   A_K^{f} ~\sim~ \sum \,g_i^{(A^f)}\,(\chi_{\sigma})^{D_c-2}  ~,\nonumber\\
     &&B_K ~\sim~ \sum \,g_i^{(B)}\,(\chi_{1}\chi_{\sigma})^{(D_c-2)/2}
  ~,~~~~ C_K ~\sim~ \sum \,g_i^{(C)}\,(\chi_{\phi})^{D_c-2}  ~.
\label{forms}
\eeqn
Here the $g_i$ indicate various coefficients in the above linear combinations.
Thus, whereas we can immediately interpret $A_K^b$ and $A_K^f$ as
corresponding to spacetime bosonic (NS) and fermionic (Ramond)
states respectively, the interpretation
of the states in the $B_K$ and $C_K$ sectors is not as clear:
they are evidently built upon unusual vacuum states which have
no analogues in the ordinary superstring, and their spacetime
interpretations are unknown.
It turns out, however, that $A_K^b$ and $A_K^f$ are precisely equal
as functions of $\tau$, suggesting that the fractional superstring $A$-sectors
enjoy a spacetime supersymmetry;  indeed, for each value of $K$ these sectors
contain a massless $N=2$ supergravity multiplet.  Thus, for consistency,
the $B_K$- and $C_K$-sectors must also be individually
spacetime supersymmetric, and indeed we find that $B_K=C_K=0$ as well.
This suggests that we should also be able to
write each $B_K$ and $C_K$ as the difference
of matching bosonic and fermionic contributions:
\beq
     B_K ~{\mathop =^ ?}~ B_K^b - B_K^f~,~~~~~~
     C_K ~{\mathop =^ ?}~ C_K^b - C_K^f~.
\label{BCsplittings}
\eeq
Until recently, however, it has not been known
how to achieve this splitting, and this has impeded progress
in understanding these sectors.

The second remarkable feature in these partition functions
is the fact that for $K>2$, some of the coefficients $g_i^{(A^{b,f})}$ in
Eq.~(\ref{forms}) turn out to be {\it negative}.  This indicates that
the contributions of certain NS and Ramond
sectors are {\it subtracted}\/ rather than
added to their respective bosonic or fermionic Fock spaces,
or more specifically that there exists a new type
of projection between {\it different}\/ parafermionic towers of
states which has the net effect of removing
large numbers of states from the
physical spectrum.  This is the internal projection discussed
above.  It can easily be verified that despite this internal projection,
the numbers of states remaining at each mass level are still
positive, and thus it is reasonable to ask whether
these remaining states fill out the Fock space
corresponding to some new worldsheet conformal field theory.
Mathematically, this would mean that
the net expressions $A_K^{b,f}$ should themselves be interpreted as the
characters $\chi'_h$ of the highest-weight sectors of
some {\it new}\/ conformal field theory:
\beq
          A_K^{b,f} ~\equiv~ \chi'_{\rm h}~.
\label{maybecharacter}
\eeq
Determining the properties of these smaller post-projection
CFT's is of course crucial for ultimately demonstrating the consistency
of these internal projections.

%========================================================================
\setcounter{footnote}{0}
\section{Recent Developments:  The Post-Projection CFT}

We shall now summarize some of the recent progress that has been
made in determining the various properties of the effective
worldsheet CFT's which survive these
internal projections.

Given the original expressions $A_K^{b,f}$ in
Eq.~(\ref{forms}), it turns out that we can determine the
central charges, highest weights,
fusion rules, and complete set of characters of the corresponding
post-projection CFT's.
The method is relatively straightforward.
As indicated in Eq.~(\ref{maybecharacter}),
we would like to regard each expression
$A_K^{b,f}(\tau)$ as a character $\chi'_i(\tau)$
in a corresponding post-projection CFT.
These expressions are not modular-invariant by themselves, however,
and by taking modular transformations we can construct the
complete sets of characters $\lbrace \chi'_i(\tau)\rbrace$
which are eigenfunctions of $T:\,\tau \to \tau+1$
and closed under $S:\,\tau \to -1/\tau$, so that
\beq
    \chi'_i(-1/\tau) ~=~\sum_{j}\,S_{ij}\,\chi'_j(\tau)~,~~~~~~
    \chi'_i(\tau+1) ~=~\exp\bigl\lbrace 2\pi i\ell_i\bigr\rbrace
     \,\chi'_i(\tau)~
\label{STtransforms}
\eeq
where $S_{ij}$ is the $S$-mixing matrix and $\ell_i$ is a parameter
denoting the phase accrued by $\chi'_i$ under $T$.
We can then simply expand any of these characters $\chi'_i(\tau)$
as a power series in $q\equiv\exp(2\pi i\tau)$:
\beq
         \chi'_{i}(\tau) ~=~  q^{\ell_i}\,\sum_{n=0}^\infty \,a_n^{(i)}\, q^n~,
\label{qexpansion}
\eeq
whereupon the effective central charge $c_{\rm eff}$
of the post-projection CFT can be determined by analyzing the growth
in the level degeneracies $a_n^{(i)}$ as a function of $n$:
\beq
       a_n^{(i)} ~\sim~ n^{-3/4}\,\exp \left\lbrace 4\pi \sqrt{
      {{c_{\rm eff} \,n}\over 24}} \,\right\rbrace
    ~~~~~~~{\rm as}~~n\to\infty~.
\label{asymptot}
\eeq
Similarly, the complete spectrum of highest weights in this
CFT can be determined by scanning the quantities $\ell_i$ in
Eqs.~(\ref{STtransforms}) and (\ref{qexpansion}),
for the highest weight $h_i$ corresponding to a given character $\chi'_i$
is given in general by
\beq
      h_i ~=~ \ell_i ~+~ c_{\rm eff}/24~.
\label{highestweights}
\eeq
Likewise, if we interpret each of the characters $\chi'_i$ as
corresponding to a certain unique primary field $\phi_i$ of
highest weight $h_i$
in the post-projection CFT, then the fusion rules of this CFT
\beq
     \lbrack \phi_i \rbrack \,\times
     \lbrack \phi_j \rbrack ~=~
     \sum_k\,N_{ijk} \,\lbrack \phi_k \rbrack
\label{fusionrules}
\eeq
can be obtained from the matrices $S_{ij}$ in
Eq.~(\ref{STtransforms}) via the Verlinde formula
\beq
        N_{ijk} ~=~ \sum_n \, {{
      S_{in}\, S_{jn}\, S_{nk} }\over{ S_{0n} }}~.
\label{Verlindeformula}
\eeq
Here $i=0$ corresponds to the identity field (or
vacuum sector) with $h_0=0$.

Although the procedure outlined above is completely general,
numerous subtleties appear for CFT's with central charges $c\geq 1$:
in these cases there are an infinite number of primary fields,
and the characters $\chi'_i$ typically correspond
not to a single primary field with highest
weight $h_i$, but rather to all of those primary fields with highest weights
$H$ satisfying $H=h_i$ (mod 1).
The fusion rules obtained must then be interpreted
accordingly, and one requires additional quantum numbers in order to
individually distinguish each of these primary fields and its
corresponding tower of states.
Such an analysis can nevertheless be performed, however, and
the details can be found in Ref.~\cite{D}.

The result we find is as follows.
Whereas the original light-cone worldsheet CFT of the fractional superstring
is given in Eq.~(\ref{originalCFT}) with central charge $c=48/(K+2)$,
we find that in the $A$-sectors
the internal projections effectively reduce this theory
down to one with $c_{\rm eff}=24/K$:
\beq
  K\geq 2:~~~~~~{\rm new~CFT}~=~
     \left\lbrace \mathop\otimes_{\mu=1}^{D_c-2=16/K}  X^\mu \right\rbrace
          ~~\otimes~~
   \left\lbrace \left( c={8\over K}\right)~{\rm theory}\right\rbrace~;
\label{finalCFT}
\eeq
moreover, this $c=8/K$ component theory surprisingly turns out
to be completely isomorphic to a
tensor product of $8/K$ bosons compactified on circles of radius
$R=1$.\footnote{
  Note that a single $R=1$ boson can be fermionized, yielding two copies
  of the $c=1/2$ Ising model.  Thus, in the $K=16$ case, this
  ``tensor product of $8/K$ bosons'' refers to the Ising model.
  Likewise, for the $K=2$ case of the ordinary superstring,
  there is no internal projection:  the initial and final central charges are
  equal, and since each $Z_2$ ``parafermion'' theory in
  Eq.~(\ref{originalCFT}) is nothing but the Ising model, the CFT's in
  Eqs.~(\ref{originalCFT}) and (\ref{finalCFT}) are indeed equivalent.}
Specifically, this means that for each relevant value of $K\geq 2$, the
$c=8/K$ post-projection theories in Eq.~(\ref{finalCFT})
have the same central charges, highest weights, fusion rules,
and characters as those of $8/K$ free compactified bosons ---
even though (as we shall discuss below) these post-projection CFT's cannot
ultimately be represented in this simple manner as a tensor product
of free bosonic worldsheet fields for $K>2$.
This close relationship between the $A$-sectors
of our fractional superstrings and free-boson theories implies
that the characters $A_K^{b,f}$, which are originally obtained
as {\it differences}\/ of parafermionic string functions
as in Eq.~(\ref{forms}),
should also be expressible directly in terms of ordinary Dedekind
$\eta$-functions and Jacobi $\vartheta$-functions, and indeed we
find \cite{ADT,AD,D}
\beq
          A_K^{b,f}(\tau) ~=~
            (D_c -2) \, \left[
      {{\vartheta_2(\tau)}\over{2\,\eta^3(\tau)}} \right]^{(D_c-2)/2}~.
\label{Atheta}
\eeq

Given these results for the $A$-sectors, it turns out that we can
make similar progress for the $B$- and $C$-sectors.  Recall that the
stumbling block for these sectors in Eq.~(\ref{BCsplittings})
was the fact that we had no guidance
as to how these expressions were to be separated into their separate
bosonic and fermionic contributions.
However, if these sectors are to be consistent with the $A$-sectors,
then their individual bosonic and fermionic components
must also experience analogous internal projections
which reduce their effective central charges from $c=48/(K+2)$ to
$c_{\rm eff}=24/K$.  We thus simply demand a splitting as
in Eq.~(\ref{BCsplittings}) such that when the individual
components $B_K^{b,f}$ and $C_K^{b,f}$ are $q$-expanded as
in Eq.~(\ref{qexpansion}),
their level degeneracies each grow as in Eq.~(\ref{asymptot})
with this value of $c_{\rm eff}$.  It turns out that this yields
a unique splitting in each case \cite{CR,AD}, and then by following the
procedure outlined above we can determine the effective post-projection
CFT for each of these sectors as well.

We find the following results \cite{D}.
For the $B$-sectors, we find that the internal projections
also reduce our original CFT down to the $c_{\rm eff}=24/K$ theory given
in Eq.~(\ref{finalCFT}), but now the $c=8/K$ component theory
in Eq.~(\ref{finalCFT}) is isomorphic to a tensor product of $8/K$ bosons
compactified on circles of radius $R=\sqrt{\lambda}$, where
\beq
        \lambda ~\equiv ~
    \half\,(K+2) ~=~ \cases{  3 & for $K=4$ \cr 5 & for $K=8$~.\cr}
\label{lambdadef}
\eeq
Indeed, in analogy to Eq.~(\ref{Atheta}),
the individual expressions $B_K^{b,f}$ can now be related
to Jacobi $\vartheta$-functions with {\it scaled}\/ arguments \cite{AD,D}:
\beq
          B_K^{b,f}(\tau) ~=~ (D_c -2) \, \left[
      {{\vartheta_2(\lambda\tau)}\over{2\,\eta^3(\tau)}}
       \right]^{(D_c-2)/2}~.
\label{Btheta}
\eeq
Note that while the internal projections in the $A$-sectors
seem to remove all traces of our original worldsheet $\IZ_K$ parafermion
theory, this $B$-sector scaling factor $\lambda$ is in fact the inverse of
the spin of the original parafermion $\epsilon$ for each value of $K$.
For the $C$-sectors, on the other hand, we find a somewhat different
story:  the only splitting consistent with the internal projections
yields expressions $C_K^{b,f}$ which each separately vanish.
Thus the internal projections actually remove {\it all}\/ $C$-sector
states from the physical spectrum, and the $C$-sectors play no
role in the post-projection worldsheet CFT.

Taken together, then, these results suggest that the internal projections
act in an internally consistent manner, with the surviving states
recombining to precisely fill out all of the momentum and
winding-mode sectors appropriate to compactified-boson worldsheet theories.
Indeed, the only difference between the $A$-sectors and the $B$-sectors
is an apparent change in the compactification radius of these
isomorphic bosons, and this indicates that although the $B$-sectors
appear very different from the $A$-sectors from
the {\it pre}\/-projection (or parafermionic)
point of view, they turn out to closely resemble
the $A$-sectors {\it after}\/ the internal projections have acted.
This isomorphism between the post-projection CFT's
and the compactified-boson theories does not imply,
however, that the former can ultimately be {\it represented}\/
in this manner --- {\it i.e.}, in terms of free bosonic
worldsheet fields.  Indeed, as we shall now discuss,
such a simple representation would not yield the correct
spacetime statistics properties for the various sectors
of our post-projection CFT's.

%========================================================================
\setcounter{footnote}{0}
\section{Lattices and Spacetime Statistics}

This isomorphism between the post-projection theories and the
compactified-boson theories enables us to go one step further,
in fact,
and actually examine the individual states which comprise
the various post-projections sectors of the fractional superstring.
This occurs because the compactified-boson theories furnish
us with an additional quantum number --- namely the $U(1)$ charge
$\alpha$ --- according to which the infinite numbers of primary
fields in these $c\geq 1$ theories may be distinguished and placed
on a lattice.
As we shall see, this proves to be of great importance in
describing the spacetime statistics of the various surviving states,
and thereby demonstrating that a simple free-boson representation
of our post-projection CFT's is unsuitable for $K>2$.

Let us first recall some features of the (chiral) compactified-boson CFT.
This theory contains primary fields $e^{i\alpha \phi}$ of
conformal dimensions $\alpha^2/2$, where
$\phi(z)$ indicates the boson field and where $\alpha$ turns out
to be the charge of the
primary field with respect to the $U(1)$ current $i\partial \phi$.
This charge is conserved under fusion:
\beq
    \lbrack e^{i\alpha\phi} \rbrack ~\times~
    \lbrack e^{i\beta\phi} \rbrack ~=~
    \lbrack e^{i(\alpha+\beta)\phi} \rbrack ~.
\label{bosonfusions}
\eeq
If $\phi$ is compactified on a circle of radius $R$, so that
$\phi\approx \phi+2\pi R$, then $\alpha$ is restricted to
the values $n/(2R)$ with $n\in\IZ$.  Thus the set of
allowed $\alpha$-values forms a one-dimensional lattice with lattice
spacing $1/2R$.  Each lattice site corresponds to a
different vacuum state $e^{i\alpha\phi(0)}|0\rangle$, and gives rise to
an infinite tower of states reached by bosonic mode excitations.
Thus, a tensor product of $8/K$ chiral bosons consists of
states $\vec\alpha$ which fill out an $8/K$-dimensional lattice:  the
corresponding highest weights in the full $8/K$-boson theory
are given by $h=\vec\alpha\cdot\vec\alpha/2$,
and its fusion rules are equivalent to vector addition
for $\vec\alpha$.
Such a (left-moving) $8/K$-dimensional lattice $\Lambda_L$ must of course
be tensored with a corresponding (right-moving) $8/K$-dimensional lattice
$\Lambda_R$ in order to fully describe
the spectrum of states in a closed string theory.

In a general $K=2$ superstring or heterotic string,
not all of these potential lattice sites
$\vec\alpha=(\vec \alpha^{\,{\rm left}}|\vec\alpha^{\,{\rm right}})$
actually contribute states to the physical spectrum,
for most suffer GSO projections and only a few
sites $\vec\alpha$ remain.  Indeed, those which remain
form not a lattice but rather a ``shifted lattice'': this means that there
exists a constant ``shift vector'' $\vec S$
such that the set $\lbrace \vec\alpha -\vec S\rbrace$ forms a true
lattice, and such that the spacetime statistics of the states
in each tower $\vec\alpha$ can be determined by computing the inner product
$(\vec\alpha-\vec S)\cdot \vec S$:
\beq
        (\vec\alpha-\vec S)  \cdot \vec S ~\in~ \cases{
           \IZ &  bosonic \cr
           \IZ + 1/2 &  fermionic~. \cr}
\label{LdotS}
\eeq
For example, for the ordinary $K=2$ Type IIA superstring,
the four-dimensional shifted lattices $\Lambda_{L,R}$
of GSO-surviving Ramond and NS states are
\beq
      \Lambda_L~=~\Lambda_{R}~=~
    \biggl\lbrace n_1,n_2,n_3,n_4 \biggr\rbrace \,\oplus\,
    \biggl\lbrace n_1-\half,n_2-\half,n_3-\half,n_4-\half \biggr\rbrace
\label{latticeleft}
\eeq
with $n_i\in\IZ$ and $\sum n_i= {\rm odd}$;  the full shifted lattice is
then given by $\Lambda_2 \equiv \Lambda_L\otimes\Lambda_R$,
and the shift vector $\vec S_{K=2}$ can be taken to be
$\vec S_{K=2}=  (1,0,0,0 \,|\, 1,0,0,0)$.
This implies that the partition function of the ordinary $K=2$ superstring
can be expressed in terms of the lattice $\Lambda_2$ of
surviving states in the usual manner:
\beq
   {\cal Z}_2 ~=~  ({\rm Im}\,\tau)^{-4} ~ |\eta|^{-24}\,
     \sum_{\vec \alpha \in \Lambda_2}
    q^{(\vec \alpha^{\rm left})^2/2}
     \,\overline q^{(\vec \alpha^{\rm right})^2/2} \,
     \exp\left\lbrack  2\pi i \,(\vec\alpha -\vec S)\cdot \vec S\right\rbrack~.
\label{Zrecast}
\eeq
Note that Eq.~(\ref{LdotS}) insures that states
contribute to ${\cal Z}_2$ with the proper statistics factor $(-1)^F$.
Indeed, in either chiral (left-moving or right-moving) sector of the
theory, the NS states $\vec\alpha^b$ appear on lattice
sites with integer components $\alpha_i\in \IZ$, while the Ramond
states $\vec\alpha^f$ have half-integer components $\alpha_i\in \IZ+1/2$.
The equality of the respective numbers of these states at each
highest weight $h=\vec\alpha\cdot\vec\alpha/2$
is consistent with the spacetime supersymmetry of the $K=2$ superstring.

Remarkably, a similar situation exists for the $K$-fractional superstring.
Those states which survive the internal and GSO projections
in the $A$-sectors again fill out
an $(8/K+8/K)$-dimensional shifted lattice $\Lambda_K$:
\beqn
   K>2:~~~~~~       \Lambda_K~\equiv~
  \left\lbrace n_1\pm \half,...,n_{8/K}\pm\half \right\rbrace
      \,\otimes\,
  \left\lbrace n_1\pm \half,...,n_{8/K}\pm\half \right\rbrace~
\eeqn
where $n_i\in \IZ$ and where each sign is chosen independently;
similarly, the $B$-sector states fill out
the lattice $\sqrt{\lambda}\,\Lambda_K$.
It then turns out \cite{D} that our fractional-superstring partition functions
can be rewritten in a manner completely analogous to Eq.~(\ref{Zrecast}):
\beqn
   |A_K^b-A_K^f|^2 ~&=&~  4\,|\eta|^{-48/K}\,
     \sum_{\vec \alpha \in \Lambda_K}
    q^{(\vec \alpha^{\rm left})^2/2}
    \, \overline q^{(\vec \alpha^{\rm right})^2/2} \,
     \exp\left\lbrack  2\pi i \,(\vec\alpha -\vec S)\cdot \vec S\right\rbrack~
    \nonumber\\
   |B_K^b-B_K^f|^2 ~&=&~  4\,|\eta|^{-48/K}\,
     \sum_{\vec \alpha \in \sqrt{\lambda}\Lambda_K}
    q^{(\vec \alpha^{\rm left})^2/2}
     \,\overline q^{(\vec \alpha^{\rm right})^2/2} \,
     \exp\left\lbrack  2\pi i \,(\vec\alpha/\sqrt{\lambda} -\vec S)\cdot
     \vec S\right\rbrack\nonumber\\
\label{latticerewrite}
\eeqn
provided the shift vectors $\vec S_K$ are now taken to be
$\vec S_4=(\half,\half\,|\half,\half)$ and $\vec S_8=(\half\,|\,\half)$.
This result is in fact consistent with spacetime supersymmetry as well,
with equal numbers of spacetime bosonic and fermionic
states at each highest weight.
Indeed, Eq.~(\ref{LdotS}) now allows us to consistently identify the
spacetime statistics of the individual chiral states that survive the
internal projections \cite{D}:
\beqn
    K=8:&&~~ \cases{
   \alpha^{b}~=~ + \half\sqrt{\lambda}
    \phantom{,\pm \half\sqrt{\lambda} ))} ~~~~
      & (mod $2\sqrt{\lambda}$) \cr
     \alpha^{f}~=~ - \half\sqrt{\lambda}
    \phantom{,\pm \half\sqrt{\lambda} ))} ~~~~
      & (mod $2\sqrt{\lambda}$) \cr}\nonumber\\
    K=4:&&~~ \cases{
     \vec \alpha^{b}~=~  (\pm \half\sqrt{\lambda},\pm \half\sqrt{\lambda})
         ~~~~ & (mod $2\sqrt{\lambda}$)\cr
     \vec \alpha^{f}~=~  (\pm \half\sqrt{\lambda},
          \mp \half\sqrt{\lambda}) ~~~~& (mod $2\sqrt{\lambda}$)~,\cr} ~
\label{alphabfs}
\eeqn
where the rescaling factor $\lambda$ is understood to be equal to $1$
for the $A$-sectors.
This identification is also consistent with an alternative analysis
making use of the so-called ``twist current'' \cite{D}.

Although these lattice results show the great similarity between the
$K$-fractional superstring and the ordinary $K=2$ superstring,
they also clearly demonstrate that
we cannot ultimately represent our post-projection CFT's in terms
of free worldsheet bosons for $K>2$, or
actually associate each lattice site $\vec\alpha$ with a primary
field $e^{i\vec\alpha\cdot\vec\phi}$.
In the superstring, for example, such a representation poses no
problem, for those states with fermionic spacetime statistics
are associated with lattice sites $\vec\alpha$ with half-integer
components $\alpha_i$, and the primary fields $e^{i\phi_i/2}$
are each equivalent to a tensor product of two Ising-model spin
fields $\sigma$ which create the necessary worldsheet cuts to alter
the boundary conditions  of worldsheet fermions and produce
fermionic spacetime statistics.
For $K>2$, however, the statistics assignments in Eq.~(\ref{alphabfs})
clearly preclude any such free-boson representation,
and only the fermionic states in the $A$-sectors appear representable
in this manner.
Therefore an alternative representation for our light-cone worldsheet
theory is needed, one which is consistent not only with these
spacetime statistics assignments, but more generally with
transverse $(D_c-2=16/K)$-dimensional Lorentz invariance.
Such issues are discussed further in Ref.~\cite{D}.

Thus, the above new results
concerning the post-projection worldsheet conformal field
theories of the fractional superstring constitute only the first steps in
their eventual construction, and many issues remain to be resolved before the
consistency of the fractional superstring is demonstrated.
Work in all of these areas is continuing.

%=============================================================================
\setcounter{footnote}{0}
\bigskip
\medskip
\leftline{\large\bf Acknowledgments}
\medskip

I am pleased to thank Philip Argyres and Henry Tye
for many useful discussions.
This work was supported in part by NSERC (Canada) and FCAR (Qu\'ebec).

%==============================================================================

\vfill\eject

\bigskip
\bibliographystyle{unsrt}

\bigskip
\noindent \underbar{\bf Note}:  Listed above are only those references
concerning fractional superstrings:
Refs.~[1--8,11] deal with the so-called ``tensor product formulation''
which has been the focus of this review, while Refs.~[9,10]
deal with an alternative ``chiral algebra formulation''.
A summary of the possible relation between the two
can be found in Ref.~[7].

\end{document}